\documentclass[aip,cha,amsmath,amssymb,reprint]{revtex4-2}
\usepackage{amsmath,amssymb,amsthm}

\theoremstyle{plain}
\newtheorem{theorem}{Theorem}

\theoremstyle{definition}

\theoremstyle{remark}

\usepackage[T1]{fontenc}
\usepackage[utf8]{inputenc}
\usepackage{lmodern}
\usepackage{microtype}
\usepackage{bm}
\usepackage{mathtools}
\usepackage{graphicx}
\usepackage{xcolor}
\usepackage{booktabs}
\usepackage{subfigure}
\usepackage{siunitx}
\usepackage{hyperref}
\usepackage[capitalize,noabbrev]{cleveref}

\hypersetup{colorlinks=true,linkcolor=blue,citecolor=blue,urlcolor=blue}

\newcommand{\vx}{\mathbf{x}}
\newcommand{\vy}{\mathbf{y}}
\newcommand{\vz}{\mathbf{z}}
\newcommand{\vf}{\mathbf{f}}
\newcommand{\vF}{\mathbf{F}}
\newcommand{\vv}{\mathbf{v}}
\newcommand{\vb}{\mathbf{b}}
\newcommand{\vg}{\mathbf{g}}
\newcommand{\vtheta}{\boldsymbol{\theta}}
\newcommand{\va}{\mathbf{a}}
\newcommand{\mA}{\boldsymbol{A}}
\newcommand{\mI}{\boldsymbol{I}}
\newcommand{\mH}{\boldsymbol{H}}
\newcommand{\UU}{\boldsymbol{U}}
\newcommand{\GG}{\boldsymbol{G}}
\newcommand{\DD}{\boldsymbol{D}} 

\newcommand{\gradx}{\nabla_{\vx}}
\newcommand{\divx}{\nabla_{\vx} \cdot}

\newcommand{\score}{\mathbf{s}} 

\begin{document}

\title{Identifying Stochastic Dynamics from Non-Sequential Data (DyNoSeD)}

\author{Zhixin Lu}
\affiliation{Allen Institute, Seattle, WA}
\author{\L{}ukasz Ku\'smierz}
\affiliation{Allen Institute, Seattle, WA}
\author{Stefan Mihalas}
\affiliation{Allen Institute, Seattle, WA}
\affiliation{University of Washington, Seattle, WA}

\date{\today}

\begin{abstract}
Inferring stochastic dynamics from data is central across the sciences, yet in many applications only unordered, non-sequential measurements are available—often restricted to limited regions of state space—so standard time-series methods do not apply. We introduce \emph{DyNoSeD}, a first-principles framework that identifies unknown dynamical parameters from such non-sequential data by minimizing Fokker–Planck residuals. We develop two complementary routes: a \emph{local} route that handles region-restricted data via locally estimated scores, and a \emph{global} route that fits dynamics from globally sampled data using a kernel Stein discrepancy without explicit density or score estimation. When the dynamics are affine in the unknown parameters \(\vtheta\) (while remaining nonlinear in the state \(\vx\)), we prove a necessary-and-sufficient condition for the \emph{existence and uniqueness} of the inferred parameter vector and derive a sensitivity analysis that identifies which parameters are tightly constrained by the data and which remain effectively free under over-parameterization. For general non-affine parameterizations, both routes define differentiable losses amenable to gradient-based optimization. As demonstrations, we recover (i) the three parameters of a stochastic Lorenz system from non-sequential observations (region-restricted data for the local route and full steady-state data for the global route) and (ii) a \(3\times 7\) interaction matrix of a nonlinear gene-regulatory network derived from a published B-cell differentiation model, using only unordered steady-state samples and applying the global route. Finally, we show that the same Fokker–Planck residual viewpoint supports a "dynamics-to-density" complement that trains a normalized density estimator directly from known dynamics without any observations. Overall, DyNoSeD provides two first-principles routes for system-identification from non-sequential data, grounded in the Fokker–Planck equation, that link data, density, and stochastic dynamics.
\end{abstract}

\maketitle

\section{Introduction}
Inferring governing dynamics from data is a central problem in science and engineering, known broadly as \emph{system-identification}~\cite{ljung1998system,isermann2011identification,brunton2016discovering,schaeffer2017learning,champion2019data,duriez2017machine}. When full time series are available, a variety of approaches---ranging from classical parametric identification~\cite{ljung1998system} to modern sparse-regression frameworks such as SINDy~\cite{brunton2016discovering}---enable the estimation of governing equations directly from observations. Recent developments extend these ideas to high-dimensional, nonlinear, and partially observed systems using machine learning and neural differential equations~\cite{raissi2019physics,chen2018neural,rackauckas2020universal}.

When continual measurements are infeasible, one may instead leverage \emph{cross-sectional} data collected at distinct time points. Some approaches construct pseudo--time series by linking samples across time points~\cite{huang2013exploiting}, while more recent work casts the problem as dynamical optimal transport over Wasserstein geodesics~\cite{tong2020trajectorynet,bunne2024optimal}. Related efforts derive estimators from the Fokker-Planck or probability-flow ODE perspectives for such cross-sectional settings~\cite{maddu2024inferring,chardes2023stochastic,chen2018neural}.

Here we study a more challenging and practically common regime in which \emph{temporal information is absent}. Data consist only of unordered \emph{steady-state measurements}\footnote{The observations are drawn after a transient phase such that the probability density of the state is time-invariant, but may still break detailed balance; individual states continue to evolve in time.} collected after the system has reached a (possibly nonequilibrium) stationary distribution. For such problems, standard time-series methods are inapplicable, and naive attempts to recover dynamics from the stationary density are typically underdetermined: many different drifts can induce the same steady law (e.g., by adding divergence-free probability currents). A central question is therefore under what conditions non-sequential steady-state data suffice to identify the underlying stochastic dynamics.

In practice, non-sequential data are often available in two distinct regimes. In some experiments, measurements can be densely curated in selected regions of state space (e.g., certain experimentally accessible ranges), but are unavailable elsewhere; here global density estimation is impossible, while local behavior is well constrained. In other settings, data are sampled unbiasedly across state space but are too sparse to support reliable global density or score estimation without imposing strong modeling biases. Our goal is to learn the dynamical parameters in \emph{both} regimes from the same first-principles starting point, and to make explicit when the resulting system-identification problem is well posed.

\begin{figure}[ht]
\centering
\includegraphics[width=0.5\columnwidth]{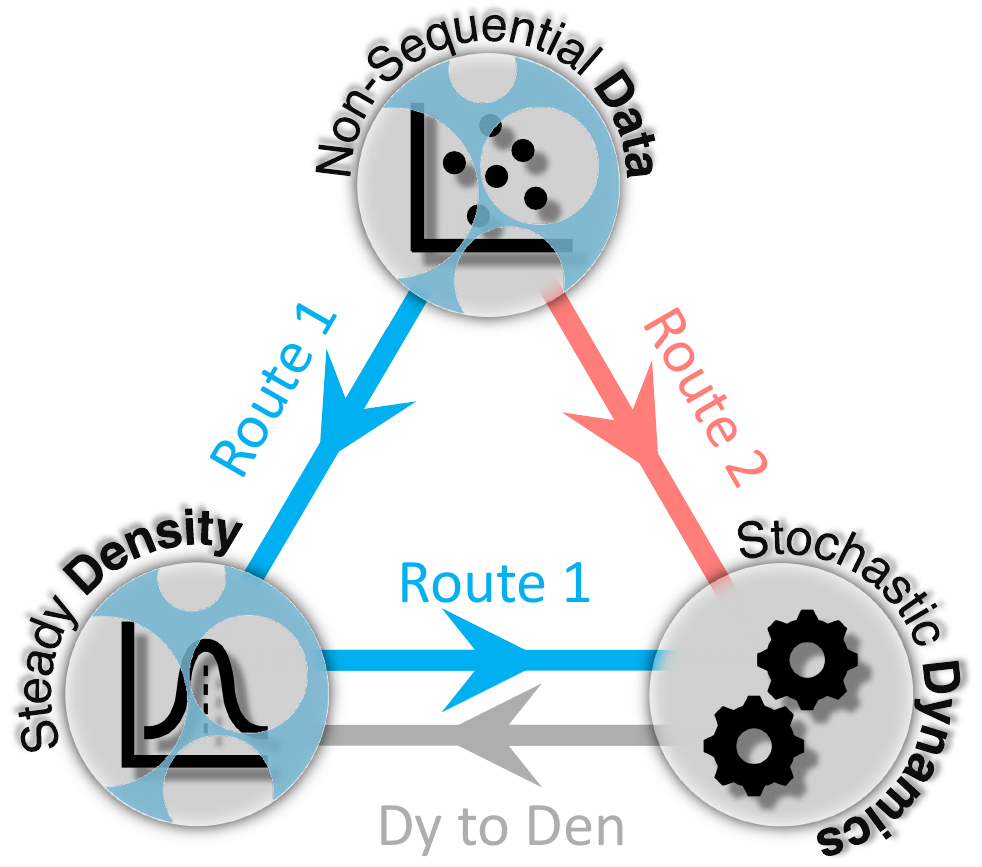}
\caption{A framework linking \emph{non-sequential data}, \emph{steady-state distributions}, and \emph{stochastic dynamics} via Fokker-Planck residuals (FPRs).
\textbf{Data$\to$Dynamics (score-based; blue):} infer dynamical parameters from unordered data---even with sampling restricted to subregions---using locally estimated scores at probe points; we provide a linear identifiability condition and first-order uncertainty analysis for affine-in-parameter priors.
\textbf{Data$\to$Dynamics (kernel Stein discrepancy; red):} infer parameters directly from broadly distributed steady-state samples without estimating densities or scores, via a kernel Stein discrepancy derived from the same FPRs; we provide a linear identifiability condition for affine-in-parameter priors.
\textbf{Dynamics$\to$Density (gray):} as a side demonstration, we use the same FPRs to infer the steady-state density directly from known dynamics.}
\label{fig:D3}
\end{figure}

We tackle this problem with a first-principles framework, \emph{DyNoSeD} (Identifying Dynamics from Non-Sequential Data), grounded in the Fokker–Planck (FP) equation. From the FP residual, we derive two complementary learning routes tailored to these two regimes (the blue and red arrows in Fig.~\ref{fig:D3}):
\begin{itemize}
  \item \textbf{Local route (score-based; blue).} When data can be densely curated in restricted regions, we infer the dynamical parameters by minimizing the Fokker-Planck residuals (FPRs) using locally estimated scores $\score(\vx)=\nabla_{\vx}\log p(\vx)$ at probe locations (e.g., simple kernel estimations, score matching\cite{hyvarinen2005estimation}, or the sliced score matching\cite{song2020sliced} that is efficient for high-dimensional data). This route never requires reconstructing the global density; it only needs an accurate local structure where data are abundant.
  \item \textbf{Global route (Stein-based; red).} When samples are broadly distributed but not dense enough to reliably estimate a global density or score, we avoid density/score estimation altogether and instead minimize the same FP residual in a global sense via a kernel Stein discrepancy (KSD). Here the kernel is used to define a universal reproducing-kernel Hilbert space whose test functions collectively enforce the vanishing of the residual. Using random Fourier features, we obtain a linear complexity KSD estimator that fits dynamical parameters directly from data without any explicit density or score model.
\end{itemize}

The DyNoSeD framework allows us to derive an explicit condition under which the unknown parameters can be uniquely determined from the available data. Specifically, when the prior dynamics are \emph{affine} in their unknown parameters~$\vtheta$ (while remaining nonlinear in the state~$\vx$), both routes share a common algebraic structure: minimizing the FPRs yields a linear system, $\mA\vtheta = \vb$, evaluated at probe points (local route) or via global averages (KSD route). Beyond identifiability, we also derive a parameter-wise sensitivity analysis for the affine case that reveals which components of \(\vtheta\) are tightly constrained by the data and which directions remain effectively free under over-parameterization. When the dynamics are not affine in $\vtheta$, both routes naturally define differentiable loss functions amenable to gradient-based optimization (e.g., with automatic differentiation), while retaining the advantages of local score estimation or linear complexity KSD evaluation.

We illustrate DyNoSeD on two canonical yet challenging systems. For a stochastic Lorenz SDE, we recover its three parameters from non-sequential data (region-restricted data via the local route and globally sampled steady-state data via the global route). For a nonlinear gene-regulatory network derived from a B-cell differentiation model~\citep{martinez2012quantitative}, we infer the \(3\times 7\) interaction matrix from unordered steady-state samples using the global route and quantify how tightly each inferred interaction is constrained.

Although our main focus is the steady-state setting, the same construction also extends to \emph{nonstationary} data. When all data are collected at a single time \(t\) and the time derivative \(\partial_t \log p(\vx,t)\) is available, the Fokker–Planck residual acquires an additional known term, and the resulting identification problem retains the same structure. We provide this extension in the SM. As a further complement (the gray arrow in Fig.~\ref{fig:D3}), we show that the same FPR can be used in the opposite direction: given known dynamics, one can train a normalized density estimator directly from the governing equations without any data. We illustrate this dynamics$\to$density route with a simple two-dimensional example.

In summary, our contributions are:
\begin{enumerate}
  \item A Fokker–Planck-based formulation of system identification from non-sequential steady-state data, with two complementary routes: a local score-based method tailored to region-restricted, locally dense sampling, and a global KSD method tailored to globally sampled data with linear complexity.
  \item A unified identifiability result for affine-in-parameter dynamical priors, in which both routes reduce to a linear system $\mA\vtheta=\vb$, together with a parameterwise sensitivity analysis based on the (regularized) Gram matrix $\mH_\lambda = \mA^\top\mA + \lambda \mI$ that reveals which parameters are well constrained by the data.
  \item Gradient-based extensions of both routes for general non-affine parameterizations.
  \item Demonstrations on a stochastic Lorenz system and a nonlinear gene-regulatory network with higher-order interactions, plus a small "dynamics-to-density" example, all using the same FP-residual viewpoint.
\end{enumerate}
Together, these elements provide two first-principles routes for system-identification from non-sequential data, grounded in the FP equation, that link data, steady-state distributions, and stochastic dynamics.

\textbf{Related work.}
Classical system-identification from time series is mature, and sparse-regression approaches such as SINDy provide scalable priors for discovering governing equations~\cite{brunton2016discovering}. When cross-sectional measurements at multiple time points are available, pseudo--time construction links samples across time~\cite{huang2013exploiting}, and dynamical optimal transport formulates learning as time-indexed flows on Wasserstein space~\cite{tong2020trajectorynet,bunne2024optimal}. Other methods leverage the FP and probability-flow viewpoints to recover dynamics from cross-sectional data~\cite{maddu2024inferring,chardes2023stochastic,chen2018neural}. Our formulation departs in two directions: (i) it targets \emph{non-sequential} data settings without cross-sectional time labels, and (ii) it offers both a \emph{local} score-based route and a \emph{global} Stein-based route derived from the same FP structure. The local route exploits the fact that scores can be estimated from unordered measurements using score matching and its variants~\cite{hyvarinen2005estimation,song2020score,song2020sliced}, enabling uneven, region-restricted sampling and yielding an explicit linear identifiability condition and associated sensitivity analysis. The global route connects to Stein discrepancies~\cite{chwialkowski2016kernel,liu2016stein,barp2019minimum}, providing a likelihood-free alternative that avoids explicit density or score estimation while inheriting the same affine-in-parameter identifiability structure.

\section{Problem setup}
\label{sec:setup_fp}

Consider a dynamical system governed by the It\^{o} stochastic differential equation (SDE)
\begin{equation}
\mathrm d\vx = \vf_{\vtheta}(\vx)\,\mathrm dt + \GG(\vx)\,\mathrm d\boldsymbol{w}_t,
\label{eqn:dyn}
\end{equation}
where $\vx\in\mathbb{R}^d$ denotes the state, $\boldsymbol{w}_t\in\mathbb{R}^{d'}$ is a standard Wiener process, $\GG(\vx)$ is a known $d\times d'$ matrix, and $\vtheta\in\mathbb{R}^n$ are unknown parameters of the drift $\vf_{\vtheta}$. The diffusion coefficient is then a known positive semidefinite matrix function
\begin{equation}
\DD(\vx) \;\equiv\; \tfrac{1}{2}\GG(\vx)\GG(\vx)^{\!\top}.
\end{equation}
For clarity of exposition in the main text, we assume a \emph{constant} diffusion $\DD$; the state-dependent case simply adds known divergence terms in~$\DD$ and can be handled analogously (see SM).

The goal is to identify the dynamical parameters $\vtheta$ from non-sequential data. We assume that, after a transient, the SDE admits a (possibly non-equilibrium) stationary density $p(\vx)$, and we observe post-transient states $\{\vx_i\}_{i=1}^N$ without time stamps. We focus on two practically common regimes: (i) samples are concentrated in several subregions of state space with possibly biased sampling rates across regions (only local information about $p$ is available there); and (ii) samples are broadly distributed so that $\{\vx_i\}_{i=1}^N$ approximate draws from $p(\vx)$, but may still be too sparse in some regions for reliable density or score estimation. The local and global routes proposed in this work are designed for these two regimes, respectively.

\begin{figure}[t]
\centering
\includegraphics[width=1\columnwidth]{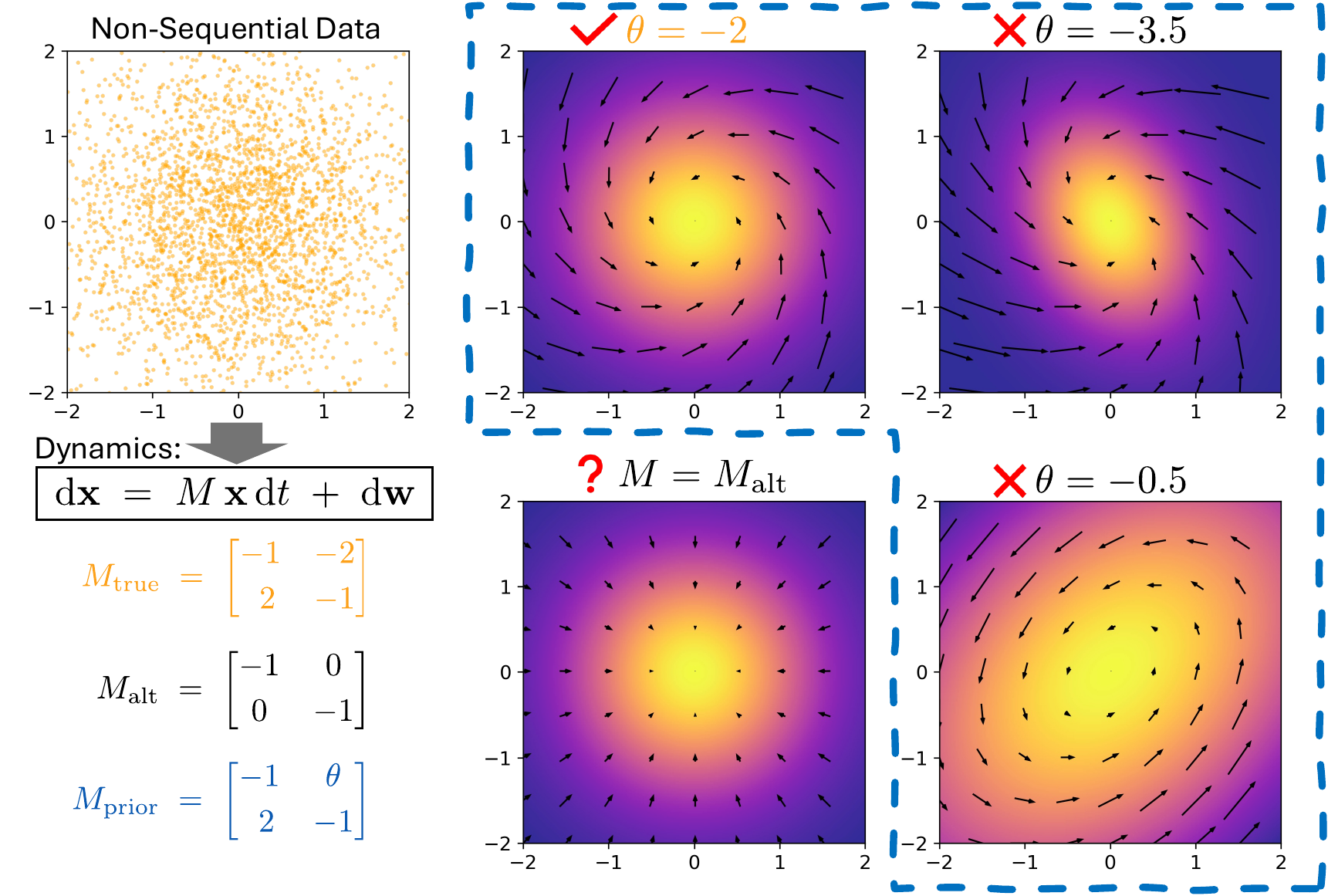}
\caption{Ill-posedness without a constraining prior (Ornstein--Uhlenbeck example). Ground-truth drift \(M_{\mathrm{true}}\) (left) and its steady density (center) admit alternative drifts with the same steady density when divergence-free probability currents are allowed. A naive norm penalty would select a flux-free diagonal \(M_{\mathrm{alt}}\) that matches the density but yields incorrect dynamics. Restricting the unknowns via an informative prior (e.g., only \(M_{12}\) free) restores identifiability and recovers the true flow. Vector fields are overlaid with level sets of the steady density.}
\label{fig:illpose}
\end{figure}

Without appropriate prior structure, learning $\vtheta$ from such non-sequential data is generically under-determined. Even for the well-understood linear Ornstein--Uhlenbeck process (Fig.~\ref{fig:illpose}), it is impossible to uniquely determine the drift $M\vx$ from the stationary density alone; many distinct drifts can generate the same stationary law by differing only in a divergence-free probability current. In more general nonlinear settings, identifiability becomes even more elusive due to the lack of global knowledge of the density function and the possible degeneracies in the parameterization of the drift $\vf_{\vtheta}$ under the given density function.

To make both estimation and identifiability tractable, we focus--when studying identifiability--on a practically common and analytically convenient class of priors in which the \(\vf_{\vtheta}\) is \emph{affine-in-parameter}:
\begin{equation}
\vf_{\vtheta}(\vx) = \UU(\vx)\vtheta + \vv(\vx),
\label{eqn:affine_drift}
\end{equation}
where $\UU:\mathbb{R}^d\to\mathbb{R}^{d\times n}$ collects $n$ nonlinear basis vector fields (columns) encoding prior knowledge, and $\vv:\mathbb{R}^d\to\mathbb{R}^d$ is a known vector field. Such affine-in-parameter priors can be highly nonlinear in the state \(\vx\); yet, as we show in the next section, they convert the identification of \(\vtheta\) into a linear system of the form $\mA\vtheta=\vb$, with a clear necessary and sufficient condition characterizing existence and uniqueness. When no such affine-in-parameters prior is available, the methods proposed in this work can still learn $\vtheta$ via gradient-based optimization.

\section{System identification via the Fokker-Planck equation}
\label{sec:fp_residual}

For the SDE~\eqref{eqn:dyn} with constant diffusion $\DD$, the density
$p(\vx,t)$ evolves according to the Fokker--Planck equation
\begin{equation}
\frac{\partial p(\vx,t)}{\partial t}
= \nabla_{\vx}\!\cdot\!\Big(-\vf_{\vtheta}(\vx)\,p(\vx,t)
  +\DD\,\nabla_{\vx} p(\vx,t)\Big).
\label{eq:fp_time}
\end{equation}
In the stationary regime of interest, $\partial_t p(\vx,t)=0$, and we
define the \emph{Fokker--Planck residual} (FPR)
\begin{equation}
R(\vx;\vtheta)
:= \nabla_{\vx}\!\cdot\!\Big(\vf_{\vtheta}(\vx)\,p(\vx)
  - \DD\,\nabla_{\vx} p(\vx)\Big),
\label{eqn:residual}
\end{equation}
which should vanish at the true parameters $\vtheta_\star$ for all $\vx$.
We now show how this residual leads to two complementary estimators: a local
score-based route (Sec.~\ref{subsec:route1}) and a global Stein-based route
(Sec.~\ref{subsec:route2}).

\subsection{Route 1: local score-based identification}
\label{subsec:route1}

When the sampled data are restricted to, or only dense in, subregions of
state space---with possibly uneven sampling across subregions---global
estimation of $p$ is not feasible. However, the \emph{score}
\begin{equation}
\score(\vx) \;\equiv\; \nabla_{\vx}\log p(\vx),
\end{equation}
can often be estimated \emph{locally} from such data. Dividing
\eqref{eqn:residual} by $p(\vx)$ and expressing derivatives in terms of
the score yields a scalar \emph{local} residual
\begin{equation}
\label{eq:residual_local}
\begin{aligned}
R_{\text{local}}(\vx;\vtheta)
\;:=\;&
\score(\vx)^{\!\top}\vf_{\vtheta}(\vx)
+\nabla_{\vx}\!\cdot \vf_{\vtheta}(\vx) \\
&-\score(\vx)^{\!\top}\DD\,\score(\vx)
-\nabla_{\vx}\!\cdot\big(\DD\,\score(\vx)\big).
\end{aligned}
\end{equation}
At an exact stationary solution, $R_{\text{local}}(\vx;\vtheta_\star)=0$
for all $\vx$.

To infer the unknown parameters $\vtheta$, we estimate the scores at $m$
\emph{probe locations} $\{\vx_i\}_{i=1}^{m}$ chosen in regions where the
non-sequential data $\{\vx_j\}_{j=1}^{N}$ are dense, using any
off-the-shelf score-estimation method such as score matching, sliced
score matching, or simple kernel-based estimators.\footnote{In all
experiments in this paper, we estimate $\score$ using a simple Gaussian
radial kernel estimator with a bandwidth (“temperature”) parameter~$\mathcal{T}$;
the examples show that the recovered parameters are robust over a broad
range of~$\mathcal{T}$ when the data is locally dense.} We then minimize the local loss
\begin{equation}
\label{eq:loss_local}
\boxed{\mathcal{L}_{\mathrm{local}}(\vtheta)
= \frac{1}{m}\sum_{i=1}^{m}
  \big|R_{\text{local}}(\vx_i;\vtheta)\big|^2},
\end{equation}
where $\{\vx_i\}_{i=1}^m$ are the probe locations. This loss can be
minimized by gradient-based optimizers such as Adam or SGD; when the
dynamics are affine in $\vtheta$, it reduces to a simple least-squares
problem (Sec.~\ref{subsec:affine-ident}).

\subsection{Route 2: global kernel Stein discrepancy}
\label{subsec:route2}

When samples are \emph{globally} and approximately unbiasedly drawn from the
steady distribution $p(\vx)$, but are too sparse to support accurate
density/score estimation, we enforce the FPR condition in a global
(integral) sense instead of pointwise. Specifically, we require that
\begin{equation}
\int R(\vx;\vtheta)\,\varphi(\vx)\,\mathrm d\vx = 0,
\label{eq:global-FP-int}
\end{equation}
for all sufficiently smooth test functions $\varphi$.

Using Stein’s method, the condition~\eqref{eq:global-FP-int} can be
rewritten as an expectation of a differential operator acting on
$\varphi$:
\begin{equation}
\mathbb{E}_{p}\!\big[\mathcal A^{(D)}_{\vtheta}\varphi(\vx)\big]=0,
\qquad \forall \varphi,
\label{eq:stein-identity-brief}
\end{equation}
where the diffusion--Stein operator\footnote{For clarity we present the
constant-diffusion case $\DD(\vx)\equiv D$ in the main text. The
state-dependent case $\DD(\vx)$, as well as a non-stationary extension
when $\partial_t \log p(\vx,t)$ is available at a fixed time $t$, are
treated in the Supplementary Materials.} is defined as
\begin{equation}
\mathcal A^{(D)}_{\vtheta}\varphi(\vx)
:= \vf_{\vtheta}(\vx)^{\!\top}\nabla_{\vx}\varphi(\vx)
 + \operatorname{Tr}\!\big(D\,\nabla_{\vx}^2\varphi(\vx)\big),
\label{eq:stein-op-main}
\end{equation}
where \(\nabla_{\vx}^2\varphi(\vx)\) is the Hessian of the test function. 

Instead of checking for all test functions, we consider $\varphi$ as any function within the unit ball of a universal reproducing-kernel Hilbert space (RKHS), \(\varphi\in\mathcal H(k),\,\|\varphi\|_{\mathcal H(k)}\le 1\)
$\mathcal H(k)$ with kernel $k(\vx,\vy)$, and then minimize the \emph{worst-case} violation of the Stein-identity (Eq.~\ref{eq:stein-identity-brief}) by minimizing
\begin{equation}
R_{\text{global}}(\vtheta)
:= \sup_{\varphi\in\mathcal H(k),\,\|\varphi\|_{\mathcal H(k)}\le 1}
\Big(\mathbb{E}_{\vx\sim p}\big[\mathcal A^{(D)}_{\vtheta}\varphi(\vx)\big]\Big)^2.
\label{eq:ksd-def}
\end{equation}
Based on the reproducing property, \(\varphi(\vx)\) can be rewritten as \(\langle\varphi(\cdot),k(\vx,\cdot)\rangle_{\mathcal H}\). Thus, by applying the sample mean and the differential operator to \(\varphi(\cdot)\), we obtain
\begin{equation}
\mathbb{E}_{\vx\sim p}\big[\mathcal A^{(D)}_{\vtheta}\varphi(\vx)\big] = \langle\varphi(\cdot),\mathbb{E}_{\vx\sim p}\big[\mathcal A^{(D)}_{\vtheta}k(\vx,\cdot)\big]\rangle_{\mathcal H}.
\label{eq:sdk_norm}
\end{equation}
By using the Schwarz inequality, the fact that \(\langle\varphi(\cdot),\varphi(\cdot)\rangle_{\mathcal{H}}\leq 1\), and the identity \(\langle k(x,\cdot),\, k(y,\cdot) \rangle_{\mathcal H} \;=\; k(x,y)\), we obtain 
\begin{equation}
\begin{aligned}
    R_{\text{global}}(\vtheta):= &\lVert \mathbb{E}_{\vx\sim p}\big[\mathcal A^{(D)}_{\vtheta}k(\vx,\cdot)\rVert_{\mathcal H}^2 \\
    =&\frac{1}{N^2}\sum_{i,j=1}^N\mathcal A_{\bm\theta,\vx_i}^{\,(D)}
     \mathcal A_{\bm\theta,\vx_j}^{\,(D)}k(\vx_i,\vx_j).
\end{aligned}
\label{eq:ksd}
\end{equation}

The computational complexity for obtaining \(R_{\text{global}}(\vtheta)\) directly from (\ref{eq:ksd}) is \(O(N^2)\), i.e. it is quadratic in the number of observations. In practice, we use a linear complexity method to minimize the KSD by choosing a shift-invariant Gaussian RBF
kernel and approximating it with $m$ random Fourier features via Bochner’s
theorem. Drawing frequencies
$\{\boldsymbol\omega_r\}_{r=1}^m\sim\mathcal N(\boldsymbol 0,\ell^{-2}I_d)$
and phases $\{c_r\}_{r=1}^m\sim\mathrm{Unif}[0,2\pi]$, we define
\begin{equation}
    \begin{aligned}
    z_r(\vx)
    &:= \sqrt{\tfrac{2}{m}}\cos(\boldsymbol\omega_r^{\!\top}\vx + c_r),\\
    \vz(\vx)&:=\big(z_1(\vx),\ldots,z_m(\vx)\big)^{\!\top},
    \label{eq:rff-features-main}
    \end{aligned}
\end{equation}
so that $k(\vx,\vy)\approx z(\vx)^{\!\top}z(\vy)$ for large $m$.

Applying the Stein operator~\eqref{eq:stein-op-main} to each feature
defines the \emph{Stein feature vector}
\begin{equation}
\vg(\vx;\vtheta)
:= \big(\mathcal A^{(D)}_{\vtheta} z\big)(\vx)
:= \begin{bmatrix}
\mathcal A^{(D)}_{\vtheta} z_1(\vx)\\[-2pt]
\vdots\\[-2pt]
\mathcal A^{(D)}_{\vtheta} z_m(\vx)
\end{bmatrix}
\in\mathbb R^m,
\label{eq:g-def-main}
\end{equation}
with components
\begin{equation}
\label{eq:g-cos-closed}
\begin{aligned}
\left[\vg(\vx;\vtheta)\right]_r
=&-\sqrt{\tfrac{2}{m}}\Big[
  \sin\!\big(\boldsymbol\omega_r^{\!\top}\vx + c_r\big)\,
    \boldsymbol\omega_r^{\!\top}\vf_{\vtheta}(\vx) \\
&\qquad\qquad\quad+
  \cos\!\big(\boldsymbol\omega_r^{\!\top}\vx + c_r\big)\,
    \boldsymbol\omega_r^{\!\top}D\,\boldsymbol\omega_r
\Big].
\end{aligned}
\end{equation}

The diffusion--Stein identity implies that, at the true parameters,
$\mathbb{E}_{\vx\sim p}[\vg(\vx;\vtheta_\star)]=\boldsymbol 0$. Given globally sampled
(non-sequential) data $\{\vx_i\}_{i=1}^N\sim p$, we therefore define the
\emph{global} KSD loss as the squared norm of the empirical mean Stein
feature:
\begin{equation}
\label{eq:lin-ksd-main}
\boxed{\mathcal{L}_{\mathrm{global}}(\vtheta)
:= \bigg\|
\frac{1}{N}\sum_{i=1}^N \vg(\vx_i;\vtheta)
\bigg\|_2^2}.
\end{equation}
This objective has linear complexity $O(Nm)$ in the number of samples $N$
and features $m$, and can be minimized over~$\vtheta$ using standard
gradient-based optimizers. In the affine-in-parameters case, the mean Stein
feature is linear in~$\vtheta$, and in the infinite-data limit the root
condition $\mathbb E_{\vx\sim p}[\vg(\vx;\vtheta)]=\boldsymbol 0$ again reduces to a
linear system (Sec.~\ref{subsec:affine-ident}).

\subsection{Identification condition in the affine-in-parameter case}
\label{subsec:affine-ident}

When the drift is affine in the unknown parameters \(\vtheta\) as in
Eq.~\eqref{eqn:affine_drift}, both routes induce linear systems of the form $\mA\vtheta=\vb$.

For the \emph{local} route, substituting $\vf_{\vtheta}$ from \eqref{eqn:affine_drift} 
into the local
residual~\eqref{eq:residual_local} and using an estimated score
$\widehat{\score}(\vx_i)$ at each probe $\vx_i$ yields a scalar equation
\begin{equation}
\va(\vx_i)^{\!\top}\vtheta = b(\vx_i),
\end{equation}
where
\begin{align}
\va(\vx_i)
&:= \UU(\vx_i)^{\top}\widehat{\score}(\vx_i)
   + \nabla_{\vx}\!\cdot \UU(\vx_i),\\
b(\vx_i)
&:= \widehat{\score}(\vx_i)^{\top}\DD\,\widehat{\score}(\vx_i)
   + \nabla_{\vx}\!\cdot\big(\DD\,\widehat{\score}(\vx_i)\big) \nonumber\\
&\quad - \widehat{\score}(\vx_i)^{\top}\vv(\vx_i)
        - \nabla_{\vx}\!\cdot \vv(\vx_i).
\end{align}
Stacking $m$ probes gives the linear system
\begin{equation}
\mA_{\mathrm{local}}\vtheta = \vb_{\mathrm{local}},
\label{eq:linear_local}
\end{equation}
with rows $\va(\vx_i)^{\top}$ and entries $b(\vx_i)$.

For the \emph{global} route, substituting Eq.~\ref{eqn:affine_drift} into
Eq.~\ref{eq:lin-ksd-main} yields
\begin{equation}
\label{eq:mu-nu-def}
\begin{aligned}
&\mathbb E_{\vx\sim p}\Big[\,
  \sin\!\big(\boldsymbol\omega_r^{\!\top}\vx + c_r\big)
  \boldsymbol\omega_r^{\!\top}\UU(\vx) \Big] \vtheta = \\
&
\mathbb E_{\vx\sim p}\Big[
  \sin\!\big(\boldsymbol\omega_r^{\!\top}\vx + c_r\big)\,
    \boldsymbol\omega_r^{\!\top}\vv(\vx) \\
&\qquad\qquad\qquad\qquad
+ \cos\!\big(\boldsymbol\omega_r^{\!\top}\vx + c_r\big)\,
    \boldsymbol\omega_r^{\!\top}D\,\boldsymbol\omega_r
\Big],
\end{aligned}
\end{equation}
for $r=1,\dots,m$. Stacking $m$ features gives the linear system
\begin{equation}
\mA_{\mathrm{global}}\vtheta = \vb_{\mathrm{global}},
\end{equation}
with each row of $\mA_{\text{global}}$ and each entry $\vb_{\text{global}}$ being defined in Eq.~\ref{eq:mu-nu-def}. 

In both routes, we thus obtain a linear system
\begin{equation}
\mA\vtheta=\vb,
\end{equation}
where $\mA$ and $\vb$ denote either the local or global matrices/vectors
above. The existence and uniqueness of $\vtheta$ are characterized by
a simple rank condition:

\begin{theorem}[Identification in the affine-in-parameter case]
\label{thm:ident}
Let $\mA\in\mathbb R^{M\times n}$ and $\vb\in\mathbb R^M$ be the matrix and
vector obtained from either the local score route or the global Stein
route, under exact scores (local) or infinite data (global). Then there
exists a parameter vector $\vtheta$ whose dynamics satisfy the corresponding
Fokker-Planck constraints if and only if $\vb\in\mathrm{range}(\mA)$; this
solution is unique if and only if $\mathrm{rank}(\mA)=n$.
\end{theorem}

The proof is immediate from linear algebra: existence of a solution
is equivalent to $\vb$ belonging to the column space of $\mA$,
$\mathrm{range}(\mA)$, and uniqueness requires a trivial null space,
$\ker(\mA)=\{\boldsymbol 0\}$, i.e.\ full column rank.

For the affine-in-parameter case, we could practically infer \(\vtheta\) by 
solving the regularized least-squares problem using
\begin{equation}
\hat\vtheta_\lambda
= \big(\mA^\top \mA + \lambda \mI\big)^{-1}
  \mA^\top \vb.
\end{equation}
To handle the over-parameterization, we use the (regularized) Gram matrix
\begin{equation}
\mH_\lambda := \mA^\top \mA + \lambda \mI,
\label{eq:gram}
\end{equation}
which encodes how well each of the inferred parameters in \(\hat\vtheta\) 
is constrained.

\section{Demonstrations}
\label{sec:experiments}

We illustrate DyNoSeD on two stochastic systems with very different structures, and we provide a simple demonstration on the ``dynamics-to-density'' application of the PFR.

\subsection{Stochastic Lorenz system}

\begin{figure}[ht]
\centering
\includegraphics[width=0.99\columnwidth]{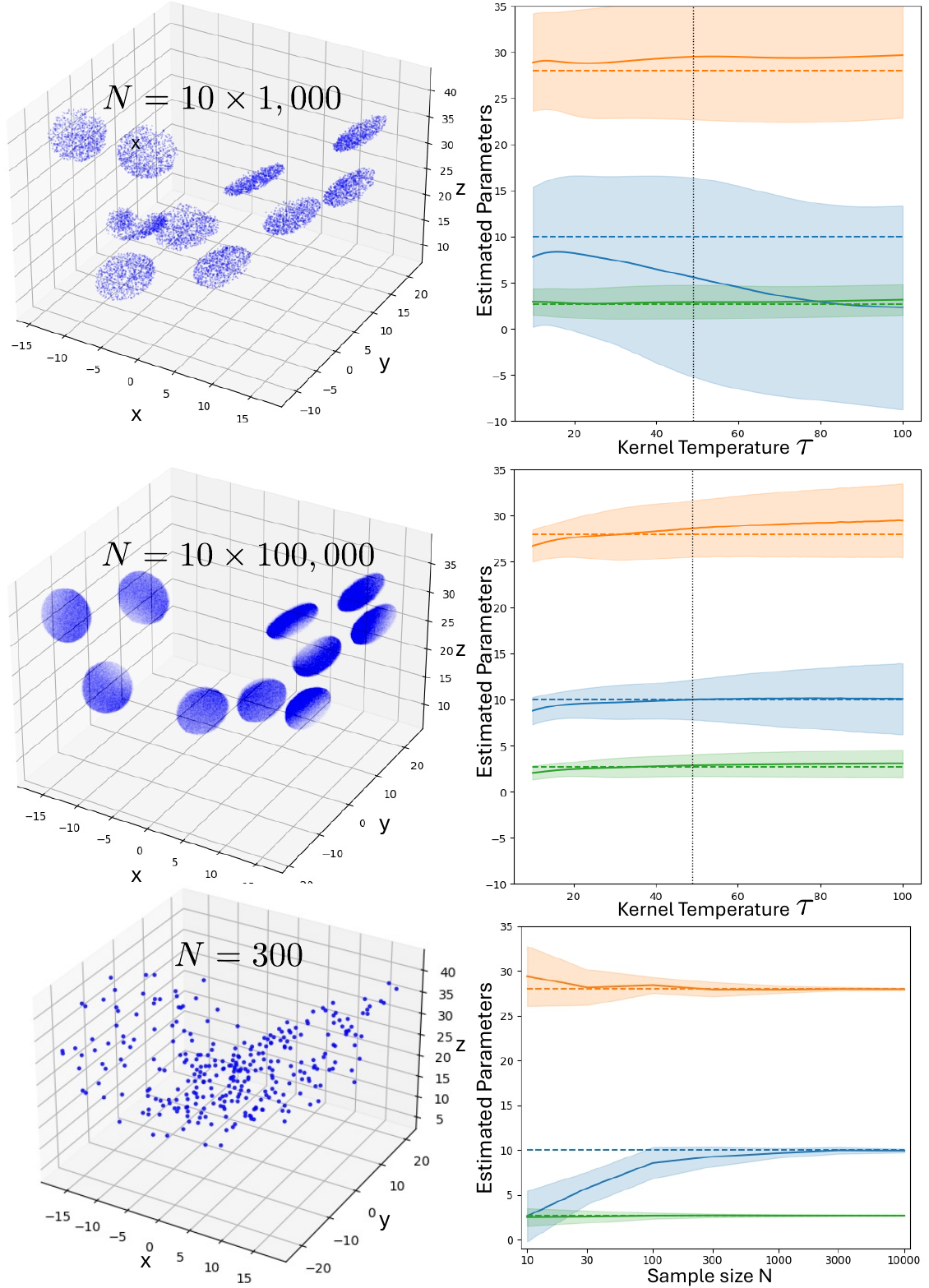}
\caption{
  \textbf{Lorenz SDE: local vs.\ global identification from non-sequential data.}
  Left column: steady-state samples on the Lorenz attractor for different sample sizes \(N\), illustrating locally dense patches (top, middle) versus a globally sparse cloud (bottom, \(N=300\)).
  Right column: recovered parameters \((\sigma,\rho,\beta)\) (solid lines: mean; shaded bands: standard deviation; dashed lines: ground truth).
  Top and middle rows: local score-based route as a function of kernel temperature \(\mathcal{T}\); estimates are accurate only when each local region is well populated.
  Bottom row: global KSD route as a function of sample size \(N\); all three parameters are recovered accurately even with a few hundred globally sampled points.
  }
  \label{fig:lorenz_local_global}
\end{figure}

We first consider the classical Lorenz SDE
\(
d\vx_t = \vf(\vx_t;\vtheta)\,dt + \sqrt{2D}\,d\mathbf{w}_t
\)
with parameters \(\vtheta=(\sigma,\rho,\beta)\) and additive isotropic noise.
We simulate long trajectories at the true parameters and thin them to obtain non-sequential steady-state samples.

For the \emph{local} route, we estimate the scores locally at the centers of the $m=10$ spheres using a Gaussian kernel with bandwidth (``temperature'') \(\mathcal{T}\).
When each box contains many points, the local route recovers all three parameters accurately over a broad range of \(\mathcal{T}\).\footnote{The quality of the score estimation depends on the Gaussian kernel bandwidth. As a reference, we denote the near–optimal bandwidth \(\mathcal{T}\approx49\) in the figure with a vertical dashed line. This value is obtained as follows. For the deterministic Lorenz system \(|\lambda_{\min}| \approx 14.57\), the most contracting Lyapunov exponent gives the relaxation rate toward the attractor. With stochastic forcing of strength $D=0.3$, the dynamics transverse to the attractor are well approximated by a one–dimensional Ornstein–Uhlenbeck mode \(dX_t = -|\lambda_{\min}| X_t\,dt + \sqrt{2D}\,dW_t\), whose stationary variance is \(D/|\lambda_{\min}|\) and hence whose characteristic precision scale is \(|\lambda_{\min}|/D\). We therefore set the kernel “temperature” to \(\mathcal{T}^\star = |\lambda_{\min}|/D \approx 14.57/0.3 \approx 49\), which empirically coincides with the bandwidth that minimizes the score estimation error.}
As the data in sphere become sparse, the estimated scores degrade, and thus the inferred parameters become strongly biased and sensitive to \(\mathcal{T}\) (top and middle rows of Fig.~\ref{fig:lorenz_local_global}).

For the \emph{global} route, we use all globally sampled data as a single cloud and minimize the linear-time KSD loss without explicit score estimation.
Even with only $N=300$ globally sampled points, the KSD route gives nearly unbiased estimates for all three parameters, and the variance shrinks rapidly with \(N\) (bottom row of Fig.~\ref{fig:lorenz_local_global}).
This highlights the complementary regimes of the two routes: local scores are powerful when data are dense in targeted regions, while the global KSD route is robust under sparse, broadly distributed sampling.

\subsection{Nonlinear gene–regulatory network}
\label{subsec:gene}
\begin{figure}[ht]
\centering
\includegraphics[width=1.0\columnwidth]{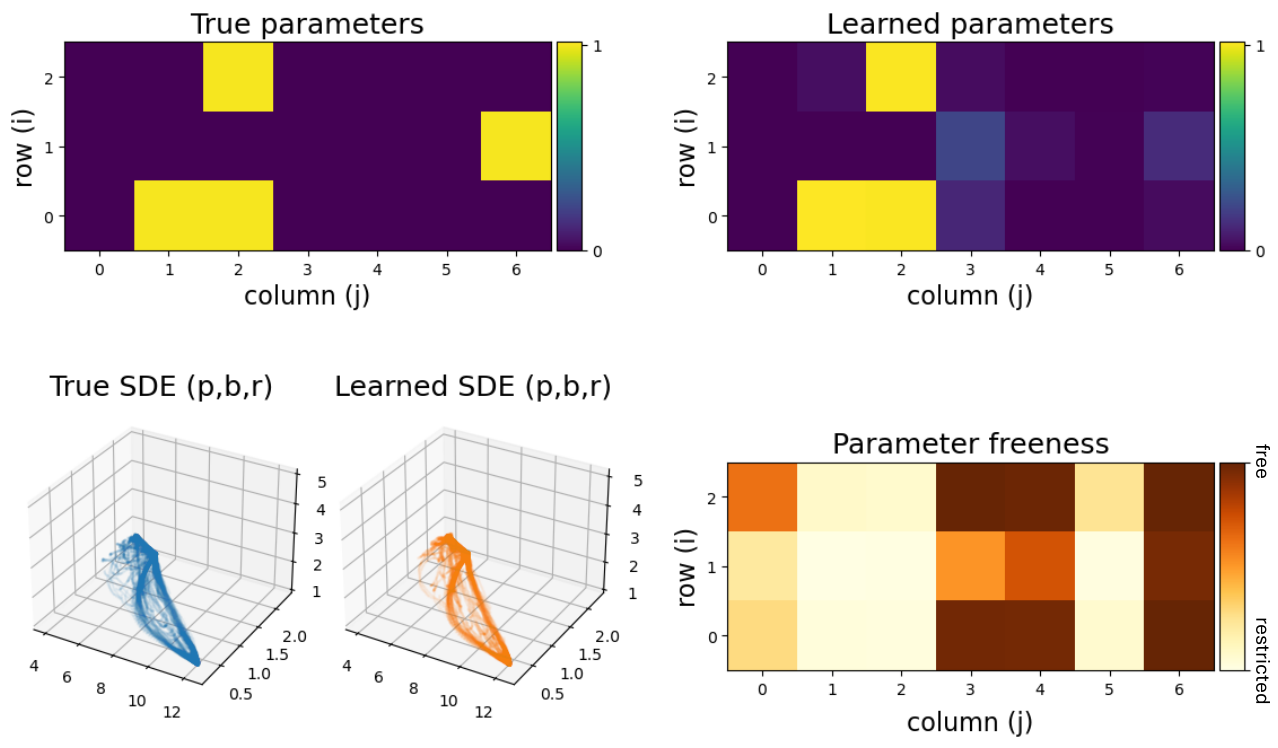}
\caption{
  \textbf{Nonlinear gene–regulatory network: parameter recovery and freeness.}
  Top row: true (left) and inferred (right) \(3\times 7\) interaction matrices for a nonlinear B-cell differentiation SDE, learned from unordered steady-state samples via the global KSD route.
  Bottom left: steady-state clouds in the \((p,b,r)\) subspace for the true (blue) and learned (orange) dynamics, which are visually indistinguishable.
  Bottom right: normalized parameter \emph{freeness} derived from the diagonal of the regularized Gram matrix \(\mH_\lambda^{-1}\); darker entries indicate directions that are less constrained by the data.
  The single badly recovered interaction coincides with a high-freeness (weakly constrained) entry.
  }
  \label{fig:grn_recover}
\end{figure}

Next we study a seven-dimensional SDE derived from a published B-cell differentiation model~\citep{martinez2012quantitative}.
Three genes \((p,b,r)\) are regulated by themselves and by two housekeeping pathways (BCR and CD40).
We encode regulation through third-order interactions of the transformed activities
\(
\pi=1/(1+p^2),\ \beta=1/(1+b^2),\ \rho=1/(1+r^2)
\), i.e., \(
\pi,\ \beta,\ \rho,\ \pi\beta,\ \pi\rho,\ \beta\rho,\ \pi\beta\rho
\), yielding a \(3\times 7\) interaction matrix.
Four additional variables describe autonomous BCR/CD40 oscillators, leading to a coupled \(7\)-dimensional SDE.
We simulate long and stochastic trajectories, thin them to obtain unordered steady-state samples, and apply the \emph{global} KSD route in its analytic affine form to recover the interaction matrix.

Figure~\ref{fig:grn_recover} (top row) compares the true and inferred \(3\times 7\) parameters. With $\lambda=10^{-6}$, most nonzero entries are recovered with small errors, but one interaction (last element in the second row) is clearly misestimated.
To understand this, we examine the Hessian matrix (Eq.~\ref{eq:gram}) and compute a parameter-wise ``freeness''
from the diagonal of \(\mH_\lambda^{-1}\); small freeness indicates a parameter is tightly constrained by the data, while large freeness indicates an effectively free parameter not constrained by the data.
The resulting heatmap (bottom-right panel) reveals that the misestimated interaction lies in one of the least constrained directions, consistent with the linear sensitivity analysis.

Crucially, the learned and true dynamics generate visually indistinguishable steady-state clouds in the \((p,b,r)\) subspace (bottom-left panels of Fig.~\ref{fig:grn_recover}), even though individual poorly constrained parameters differ.
This illustrates how DyNoSeD, together with the Gram-based sensitivity analysis, can separate parameters that are reliably identified from those that are effectively free under over-parameterization.

\subsection{``Dynamics-to-density'' complement}

\begin{figure}[ht]
\centering
\includegraphics[width=0.99\columnwidth]{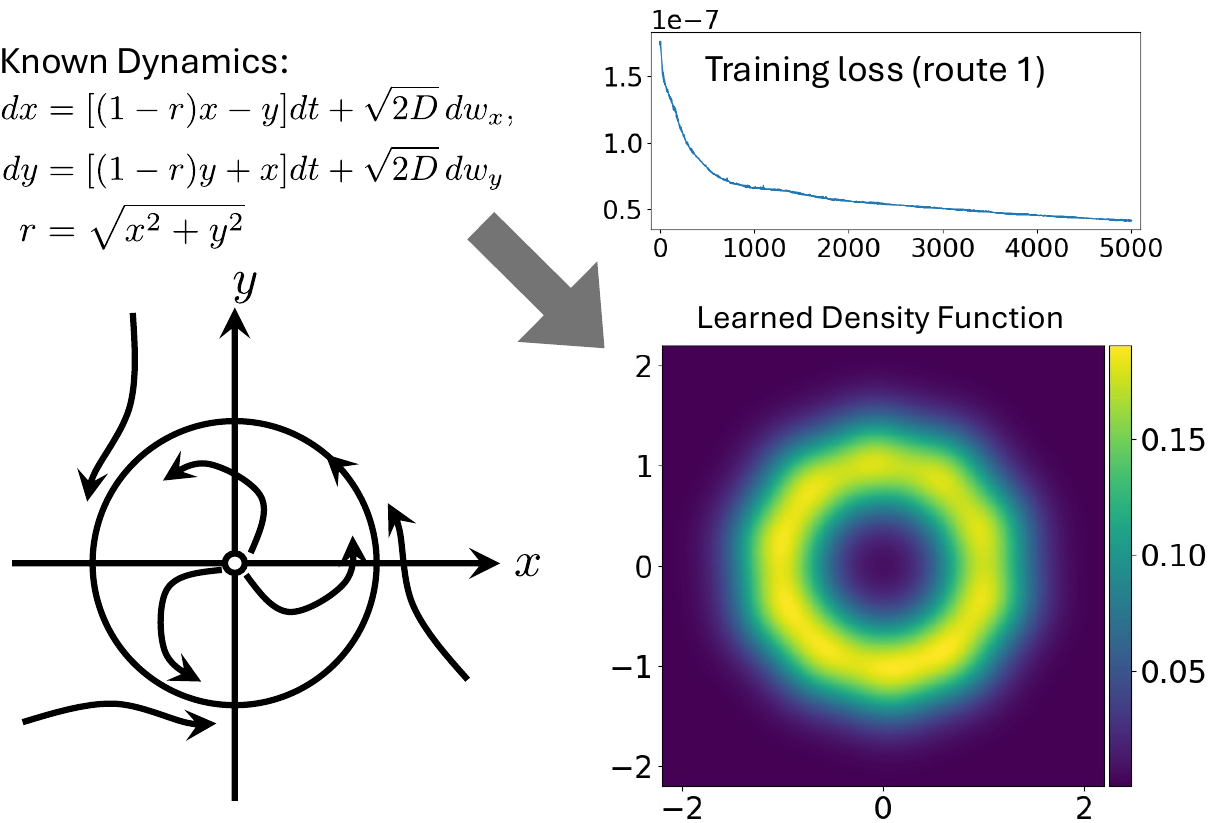}
\caption{
  \textbf{Dynamics$\to$density via Fokker–Planck residual minimization.}
  Left: schematic of a two-dimensional SDE whose drift has a stable limit cycle.
  Top right: training loss of the FP-residual objective (route 1) when fitting a neural score model \(\score_\psi(\vx)\).
  Bottom right: learned stationary density \(q_\psi\), which recovers the ring-shaped true density without using any observed data.
  }
  \label{fig:dyn2dens}
\end{figure}

Finally, we demonstrate a complementary use of the Fokker–Planck residual: given known dynamics, we train a normalized density estimator without any sampled data.
We consider a two-dimensional SDE whose underlying ODE has a stable limit cycle.
Using a neural score model \(\score_\psi(\vx)=\nabla_\vx \log q_\psi(\vx)\), we minimize the squared FP residual (route 1 as shown in Eq.~\ref{eq:loss_local}) over \(\psi\).
The training loss decreases steadily and the learned density \(q_\psi\) matches the true ring-shaped stationary density (Fig.~\ref{fig:dyn2dens}), showing that the same FPR viewpoint supports a dynamics$\to$density mapping in addition to data$\to$dynamics.

\section{Conclusion}
\label{sec:conclusion}

We introduced DyNoSeD, a Fokker–Planck–based framework for identifying stochastic dynamics from non-sequential data.
By deriving both a local score-based route and a global Stein-based route from the same FP residual, we can handle region-restricted dense sampling and globally sparse sampling within a unified formulation.
In the affine-in-parameter case, both routes reduce to a linear system \(\mA\vtheta=\vb\), yielding a simple rank-based identifiability condition and a Gram-matrix sensitivity analysis that reveals which parameters are well constrained and which are effectively free.

Our demonstrations on the Lorenz system and a nonlinear gene–regulatory network show that DyNoSeD can recover both low-dimensional and over-parameterized dynamics from unordered steady-state samples, and that the sensitivity analysis provides interpretable parameterwise reliability.
The dynamics$\to$density example further illustrates that the same FP residual can be used in reverse to learn stationary densities from known dynamics.
We expect these ideas to be useful in applications where only snapshot measurements are available, and where understanding which aspects of a mechanistic model are truly constrained by such data is as important as fitting the model itself.

\section*{Supplementary Material}
See the supplementary material for additional proofs and implementation details.
\bibliographystyle{aipnum4-2}
\bibliography{bibliography}

\clearpage
\onecolumngrid

\section*{Supplementary Material}
\addcontentsline{toc}{section}{Supplementary Material}

\setcounter{section}{0}
\renewcommand{\thesection}{S\arabic{section}}
\renewcommand{\thesubsection}{S\arabic{section}.\arabic{subsection}}
\renewcommand{\thesubsubsection}{S\arabic{section}.\arabic{subsection}.\arabic{subsubsection}}
\setcounter{equation}{0}
\renewcommand{\theequation}{S\arabic{equation}}
\setcounter{figure}{0}
\renewcommand{\thefigure}{S\arabic{figure}}

\section{Deriving Local Route for the Most General Case}
\subsection{Local route for nonstationary data}
% ...

In this section we extend the local route to the more general setting of
\emph{nonstationary} data. We consider time–varying SDEs
\begin{equation}
  d\vx_t
  =
  \vf_{\vtheta}(\vx_t,t)\,dt
  +
  \mathbf{G}(\vx_t,t)\,d\mathbf{w}_t,
  \label{eq:fukll_FPE}
\end{equation}
with state-- and time--dependent diffusion
\begin{equation}
  \DD(\vx,t)
  :=
  \tfrac12\,\mathbf{G}(\vx,t)\mathbf{G}(\vx,t)^\top.
\end{equation}
Denote the $i$--th component of $\vf_{\vtheta}$ by $f_i$. If
$p(\vx,t)$ is the time--dependent density of $\vx_t$, then $p$ obeys
the Fokker--Planck equation
\begin{align}
  \frac{\partial p(\vx,t)}{\partial t}
  &=
  - \sum_{i=1}^d \frac{\partial}{\partial x_i}
    \big[ f_i(\vx,t)\,p(\vx,t) \big]
  + \sum_{i=1}^d \sum_{j=1}^d \frac{\partial^2}{\partial x_i \partial x_j}
    \big[ \DD_{ij}(\vx,t)\,p(\vx,t) \big]
  \\
  &=
  \sum_{i=1}^d \frac{\partial}{\partial x_i}
  \Big(
    - f_i(\vx,t)\,p(\vx,t)
    + \sum_{j=1}^d \frac{\partial}{\partial x_j}
      [\DD_{ij}(\vx,t)\,p(\vx,t)]
  \Big).
  \label{eq:FP-nonstat}
\end{align}
Expanding the inner term and introducing the score
$\score(\vx,t) := \nabla_{\vx} \log p(\vx,t)$, we obtain
\begin{align}
  \sum_{j=1}^d \frac{\partial}{\partial x_j}
    [\DD_{ij}(\vx,t)\,p(\vx,t)]
  &=
  \sum_{j=1}^d
  \Big(
    \frac{\partial \DD_{ij}}{\partial x_j}(\vx,t)\,p(\vx,t)
    +
    \DD_{ij}(\vx,t)\,\frac{\partial p}{\partial x_j}(\vx,t)
  \Big)
  \\
  &=
  p(\vx,t)\,\big(\divx\DD(\vx,t)\big)_i
  +
  p(\vx,t)\,\big(\DD(\vx,t)\score(\vx,t)\big)_i,
\end{align}
where we define the matrix divergence
\[
  \big(\divx\DD(\vx,t)\big)_i
  :=
  \sum_{j=1}^d \frac{\partial \DD_{ij}}{\partial x_j}(\vx,t).
\]

In vector notation, Eq.~\eqref{eq:FP-nonstat} becomes
\begin{equation}
  \frac{\partial p(\vx,t)}{\partial t}
  =
  -\,\divx
  \Big(
    \vf_{\vtheta}(\vx,t)\,p(\vx,t)
    - p(\vx,t)\,\divx\DD(\vx,t)
    - p(\vx,t)\,\DD(\vx,t)\score(\vx,t)
  \Big).
  \label{eq:FP-nonstat-current}
\end{equation}
Introduce
\begin{equation}
  \vF_{\vtheta}^{(\DD)}(\vx,t)
  :=
  \vf_{\vtheta}(\vx,t)
  - \divx\DD(\vx,t)
  - \DD(\vx,t)\score(\vx,t),
\end{equation}
so that Eq.~\eqref{eq:FP-nonstat-current} reads
\begin{equation}
  \frac{\partial p(\vx,t)}{\partial t}
  =
  -\,\divx\big(p(\vx,t)\,\vF_{\vtheta}^{(\DD)}(\vx,t)\big).
\end{equation}
Assuming $p(\vx,t)>0$ on the region of interest, divide both sides by $p$:
\begin{equation}
  \frac{\partial}{\partial t}\log p(\vx,t)
  =
  -\,\frac{1}{p(\vx,t)}\,\divx\big(p(\vx,t)\,\vF_{\vtheta}^{(\DD)}(\vx,t)\big).
\end{equation}
Using $\gradx p = p\,\score$ and the product rule,
\[
  \divx\big( p \vF_{\vtheta}^{(\DD)} \big)
  =
  (\vF_{\vtheta}^{(\DD)})^\top \gradx p
  + p \divx\vF_{\vtheta}^{(\DD)} 
  =
  p \big(\score^\top\vF_{\vtheta}^{(\DD)}  + \divx\vF_{\vtheta}^{(\DD)} \big),
\]
we arrive at
\begin{equation}
  \frac{\partial}{\partial t}\log p(\vx,t)
  + \score(\vx,t)^\top\vF_{\vtheta}^{(\DD)}(\vx,t)
  + \divx\vF_{\vtheta}^{(\DD)}(\vx,t)
  = 0.
\end{equation}
Equivalently,
\begin{equation}
  \frac{\partial}{\partial t}\log p(\vx,t)
  + \score(\vx,t)^\top
    \big(\vf_{\vtheta} - \divx\DD - \DD\score\big)
  + \divx\big(\vf_{\vtheta} - \divx\DD - \DD\score\big)
  = 0.
\end{equation}

We define the \emph{nonstationary Fokker–Planck residual} at $(\vx,t)$ by
\begin{equation}
\begin{aligned}
    R(\vx,t;\vtheta)
  :=&
  \partial_t\log p(\vx,t) + \\
  &  \score(\vx,t)^\top
    \big(\vf_{\vtheta}(\vx,t) - \divx\DD(\vx,t) - \DD(\vx,t)\score(\vx,t)\big) + \\
  & \divx\big(\vf_{\vtheta}(\vx,t) - \divx\DD(\vx,t) - \DD(\vx,t)\score(\vx,t)\big),
\end{aligned}
  \label{eq:nonsteady_residual}
\end{equation}
so that $R(\vx,t;\vtheta^*)=0$ for the true parameters~$\vtheta^*$.

Suppose we have a collection of measurements at (possibly one or multiple) time
points,
\[
  \Omega = \{(\vx_i,t_i)\}_{i=1}^N,
\]
together with local estimates of the score $\score(\vx_i,t_i)$ and the time
derivative $\partial_t \log p(\vx_i,t_i)$ (e.g., from a parametric or neural
density model). The local nonstationary route then fits~$\vtheta$ by minimizing
the empirical FP residual, e.g.\ via
\begin{equation}
  \mathcal{L}_{\mathrm{local}}(\vtheta;\Omega)
  :=
  \frac{1}{N}\sum_{i=1}^N
    \Bigl(R(\vx_i,t_i;\vtheta)\Bigr)^2,
  \label{eq:general_loss}
\end{equation}
which reduces to the steady-state local loss when $\partial_t\log p \equiv 0$.

\section{General diffusion--Stein operator and linear–complexity KSD}
\label{apx:stein-general}

\subsection{Diffusion--Stein operator for state–dependent, non–stationary SDEs}
\label{apx:stein-operator}

Consider the time-dependent It\^{o} SDE with state– and time–dependent diffusion matrix as specified in Eq.~\ref{eq:fukll_FPE}. Let $p(\vx,t)$ denote the density of $\vx_t$.  The associated
Fokker-Planck equation is
\begin{equation}
\frac{\partial p(\vx,t)}{\partial t}
 + \nabla_{\vx}\!\cdot\!\Big(\vf_{\vtheta}(\vx,t)\,p(\vx,t)
- \divx \big(\DD(\vx,t)p(\vx,t)\big)\Big) = 0.
\label{eq:fp-general-time}
\end{equation}

Let $\varphi:\mathbb R^d\to\mathbb R$ be a smooth test function with
sufficient decay so that boundary terms vanish under integration by
parts. Multiplying~\eqref{eq:fp-general-time} by $\varphi(\vx)$ and integrating
over $\vx$ gives
\begin{equation}
    \int \Big[\frac{p(\vx,t)}{p(\vx,t)}\,
\frac{\partial p(\vx,t)}{\partial t}  + \nabla_{\vx}\!\cdot\!\Big(\vf_{\vtheta}(\vx,t)\,p(\vx,t)
-\divx \big(\DD(\vx,t)p(\vx,t)\Big)\Big]\varphi(\vx)\mathrm d\vx =0
\end{equation}

By applying integration-by-parts (i.e., the Divergence Theorem) multiple times, we could obtain 
\begin{equation}
    -\int p(\vx,t)\,\varphi(\vx)\,\partial_t\log p(\vx,t)\,\mathrm d\vx
    \;+
    \int p(\vx,t)\,\Big[
      \vf_{\vtheta}(\vx,t)\!\cdot\!\nabla_{\vx}\varphi(\vx)
      +
      \operatorname{Tr}\!\big(\DD(\vx,t)\,\nabla_{\vx}^2\varphi(\vx)\big)
    \Big]\mathrm d\vx
    = 0,
\end{equation}
which leads to 
\begin{equation}
\mathbb E_{\vx\sim p(\vx,t)}\!\big[-\big(\partial_t\log p(\vx,t)\big)\varphi(\vx) + \mathcal A^{(\DD)}_{\vtheta,t}\varphi(\vx)\big] = 0,
\label{eq:stein-evolution-general}
\end{equation}
where we have defined the diffusion--Stein operator
\begin{equation}
\mathcal A^{(\DD)}_{\vtheta,t}\varphi(\vx)
:=
\vf_{\vtheta}(\vx,t)\!\cdot\!\nabla_{\vx}\varphi(\vx)
+
\operatorname{Tr}\!\big(\DD(\vx,t)\,\nabla_{\vx}^2\varphi(\vx)\big)
.
\label{eq:stein-op-general}
\end{equation}
For constant diffusion, $\DD(\vx,t)\equiv D$, the $\mathcal A^{(\DD)}_{\vtheta,t}$ reduces to the operator $\mathcal A^{(D)}_{\vtheta}$
used in the main text (Eq.~\ref{eq:stein-op-main}). 

Two special cases are of particular interest:
\begin{itemize}
\item \textbf{Stationary regime.}  
If $p(\vx,t)$ has reached a stationary density $p^*(\vx)$, then
$\partial_t p(\vx,t)=0$ and $\partial_t\log p(\vx,t)=0$, so
\eqref{eq:stein-evolution-general} reduces to the standard
diffusion--Stein identity
\begin{equation}
\mathbb E_{\vx\sim p^*}\!\big[\mathcal A^{(\DD)}_{\vtheta}\varphi(\vx)\big]=0,
\qquad
\forall\,\varphi,
\label{eq:stein-stationary-general}
\end{equation}
with $\mathcal A^{(\DD)}_{\vtheta}:=\mathcal A^{(\DD)}_{\vtheta,t}$ evaluated at stationarity.

\item \textbf{Non–stationary snapshot.}
Fix a time $t_0$ and suppose we have access to
$\partial_t\log p(\vx,t_0)$ (e.g.\ from a density model, which does not require sequential data as it only needs the changing rate of the log likelihood at each provided data point).  Define the
augmented Stein operator
\begin{equation}
\mathcal B_{\vtheta,t_0}\varphi(\vx)
:=
- \varphi(\vx)\,\partial_t\log p(\vx,t_0)
+
\mathcal A^{(\DD)}_{\vtheta,t_0}\varphi(\vx).
\label{eq:augmented-stein-op}
\end{equation}
Then~\eqref{eq:stein-evolution-general} can be written compactly as
\begin{equation}
\mathbb E_{\vx\sim p(\vx,t_0)}\!\big[\mathcal B_{\vtheta,t_0}\varphi(\vx)\big]=0,
\qquad
\forall\,\varphi,
\label{eq:stein-identity-B}
\end{equation}
which generalizes the stationary Stein identity to a non–stationary
snapshot at $t_0$.
\end{itemize}

In the main text we focus on the stationary case with constant
diffusion, so $\mathcal B_{\vtheta,t_0}$ reduces to
$\mathcal A^{(D)}_{\vtheta}$ and~\eqref{eq:stein-identity-B} becomes
Eq.~\ref{eq:stein-identity-brief} in the main text.

\subsection{Kernel Stein discrepancy}
\label{apx:ksd-general}

Let $k(\vx,\vy)$ be a positive–definite kernel with RKHS
$\mathcal H(k)$ and reproducing property
\[
\varphi(\vx) \;=\; \langle \varphi(\cdot),\,k(\vx,\cdot)\rangle_{\mathcal H(k)}
\qquad\text{for all } \varphi\in\mathcal H(k).
\]
We then let $\mathcal B_{\vtheta,t_0}$, the (possibly nonstationary) Stein
operator for the most general case given above, be applied on to \(\varphi(\vx)\), the reproducing property yields,
\[
\mathcal B_{\vtheta,t_0}\varphi(\vx) \;=\; \langle \varphi(\cdot),\,\mathcal B_{\vtheta,t_0}k(\vx,\cdot)\rangle_{\mathcal H(k)}
\qquad\text{for all } \varphi\in\mathcal H(k).
\]
Now, we take the sample mean and obtain
\[
\mathbb{E}_{\vx\sim p(\vx,t)}[\mathcal B_{\vtheta,t_0}\varphi(\vx)] \;=\; \langle \varphi(\cdot),\,\mathbb{E}_{\vx\sim p(\vx,t)}[\mathcal B_{\vtheta,t_0}k(\vx,\cdot)]\rangle_{\mathcal H(k)}
\qquad\text{for all } \varphi\in\mathcal H(k).
\]

Now, let's only consider \(\varphi\in\mathcal{H}_k\) that is within the surface of the unit ball, i.e., \(\|\varphi\|_{\mathcal{H}}\leq1\). Then, by using the Schwarz inequality, we obtain
\begin{equation}
    \Big(\langle \varphi(\cdot),\,\mathbb{E}_{\vx\sim p(\vx,t)}[\mathcal B_{\vtheta,t_0}k(\vx,\cdot)]\rangle_{\mathcal H(k)}\Big)^2
    \leq \|\varphi(\cdot)\|^2_{\mathcal{H}}\  \|\mathbb{E}_{\vx\sim p(\vx,t)}[\mathcal B_{\vtheta,t_0}k(\vx,\cdot)]\|^2_{\mathcal{H}}.
\end{equation}
Thus, we are guaranteed that worst squared kernel Stein discrepancy in Eq.~\ref{eq:stein-identity-B} is bounded by the inequality, 
\begin{equation}
        \big(\mathbb E_{\vx\sim p(\vx,t_0)}\!\big[\mathcal B_{\vtheta,t_0}\varphi(\vx)\big] \big)^2 \;\leq\; \langle \mathbb E_{\vx\sim p(\vx,t_0)}\!\big[B_{\vtheta,t_0;\vx}k(\vx,\cdot)\big],\mathbb E_{\vy\sim p(\vy,t_0)}\!\big[B_{\vtheta,t_0;\vy}k(\vy,\cdot)\big]\rangle_{\mathcal H(k)}.
        \label{eq:ineq}
\end{equation}
By taking the sample mean out, we obtain
\begin{equation}
\langle \mathbb E_{\vx\sim p(\vx,t_0)}\!\big[B_{\vtheta,t_0;\vx}k(\vx,\cdot)\big],\mathbb E_{\vy\sim p(\vy,t_0)}\!\big[B_{\vtheta,t_0;\vy}k(\vy,\cdot)\big]\rangle_{\mathcal H(k)} = \mathbb E_{\vx,\vy}\Big[
\big\langle
\mathcal B_{\vtheta,t_0,\vx}k(\vx,\cdot),\,
\mathcal B_{\vtheta,t_0,\vy}k(\vy,\cdot)
\big\rangle_{\mathcal H(k)}
\Big],
\label{eq:average_out}
\end{equation}
where $\vx,\vy\sim p(\cdot,t_0)$ independently.
Here, we guarantee that for any fixed $\vy$, the function
$\mathcal{B}_{\vtheta,t_0,\vy}k(\vy,\cdot)$ remains in $\mathcal{H}(k)$ by choosing kernel that is universal. Then, by applying the two operators in Eq.~\ref{eq:average_out} (one on \(\vx\) and the other on \(\vy\)) onto the reproducing property,
\[
k(x,y) = \langle k(\vx,\cdot), k(\vy,\cdot) \rangle_\mathcal{H}
\]
Eqs.~\ref{eq:ineq}-\ref{eq:average_out} yields
\begin{equation}
\mathrm{KSD}^2(\vtheta)
=
\mathbb E_{\vx,\vy\sim p(\cdot,t_0)}\!
\big[k_{\vtheta}(\vx,\vy)\big],
\label{eq:ksd-kernel-expectation}
\end{equation}
with the \emph{Stein kernel}
\begin{equation}
k_{\vtheta}(\vx,\vy)
:=
\mathcal{B}_{\vtheta,t_0,\vx}\mathcal{B}_{\vtheta,t_0,\vy}
k(\vx,\vy).
\label{eq:stein-kernel-general}
\end{equation}

Given i.i.d.\ samples $\{\vx_i\}_{i=1}^N\sim p(\cdot,t_0)$, the standard
estimator of~\eqref{eq:ksd-kernel-expectation} is
\begin{equation}
\widehat{\mathrm{KSD}}^{\,2}(\vtheta)
:=
\frac{1}{N^2}
\sum_{i,j=1}^N
k_{\vtheta}(\vx_i,\vx_j),
\label{eq:ksd-vstat-general}
\end{equation}
which reduces to Eq.~\ref{eq:ksd} in the main text when
$\partial_t\log p(\vx,t_0)\equiv 0$ and $\DD(\vx,t_0)\equiv D$ is
constant. The direct computation of~\eqref{eq:ksd-vstat-general} requires
$O(N^2)$ time; in the next subsection we show how to obtain a linear complexity approximation using random Fourier features.

\subsection{Random Fourier features (RFFs) and linear–complexity KSD}
\label{apx:rff-general}

To obtain a linear–time approximation, we specialize to a
shift–invariant kernel $k(\vx,\vy)=k(\vx-\vy)$ with spectral density
$p(\boldsymbol\omega)$ and use random Fourier features.
Here, we consider the Gaussian RBF kernel
\begin{equation}
k(\vx,\vy)
=
\exp\!\Big(-\frac{\|\vx-\vy\|_2^2}{2\ell^2}\Big).
\end{equation}
Bochner’s theorem states that any continuous, positive–definite,
shift–invariant kernel admits the representation
\begin{equation}
k(\vx-\vy)
=
\int_{\mathbb R^d} e^{i\boldsymbol\omega^{\!\top}(\vx-\vy)}
\,p(\boldsymbol\omega)\,\mathrm d\boldsymbol\omega
=
\mathbb E_{\boldsymbol\omega}
\big[
e^{i(\boldsymbol\omega^{\!\top}\vx-\boldsymbol\omega^{\!\top}\vy)}
\big],
\end{equation}
where $p(\boldsymbol\omega)$ is the kernel’s spectral density
(for the RBF kernel, $p(\boldsymbol\omega)
= \mathcal N(\boldsymbol 0,\ell^{-2}I_d)$).

Taking the real part and using
$e^{i\alpha} = \cos\alpha + i\sin\alpha$ yields
\begin{align}
k(\vx,\vy)
&= \mathbb E_{\boldsymbol\omega}
\big[\cos(\boldsymbol\omega^{\!\top}\vx)\cos(\boldsymbol\omega^{\!\top}\vy)
     + \sin(\boldsymbol\omega^{\!\top}\vx)\sin(\boldsymbol\omega^{\!\top}\vy)
\big] \\
&= \mathbb E_{\boldsymbol\omega}
\Big[
\underbrace{
\begin{pmatrix}
\cos(\boldsymbol\omega^{\!\top}\vx) \\
\sin(\boldsymbol\omega^{\!\top}\vx)
\end{pmatrix}}_{:=\,\psi_{\boldsymbol\omega}(\vx)}\cdot
\underbrace{
\begin{pmatrix}
\cos(\boldsymbol\omega^{\!\top}\vy) \\
\sin(\boldsymbol\omega^{\!\top}\vy)
\end{pmatrix}}_{:=\,\psi_{\boldsymbol\omega}(\vy)}
\Big].
\end{align}
Thus one natural feature map uses both cosine and sine components
$\psi_{\boldsymbol\omega}(\vx)\in\mathbb R^2$.

To avoid carrying two trigonometric components per frequency, it is
standard to introduce a random phase $b\sim\mathrm{Unif}[0,2\pi]$ and
use a single cosine feature.  A direct computation shows that
\begin{align}
\mathbb E_{c\sim\mathrm{Unif}[0,2\pi]}
\big[2\cos(\boldsymbol\omega^{\!\top}\vx + c)\cos(\boldsymbol\omega^{\!\top}\vy + c)\big]
&=
\cos(\boldsymbol\omega^{\!\top}\vx)\cos(\boldsymbol\omega^{\!\top}\vy)
+
\sin(\boldsymbol\omega^{\!\top}\vx)\sin(\boldsymbol\omega^{\!\top}\vy),
\end{align}
because the cross–terms integrate to zero when $c$ is uniform on
$[0,2\pi]$.  Therefore,
\begin{equation}
k(\vx,\vy)
=
\mathbb E_{\boldsymbol\omega,c}
\big[
\cos(\boldsymbol\omega^{\!\top}\vx + c)\,
\cos(\boldsymbol\omega^{\!\top}\vy + c)
\big],
\qquad
\boldsymbol\omega\sim p(\boldsymbol\omega),\;
c\sim\mathrm{Unif}[0,2\pi].
\end{equation}
The sine terms are thus “hidden” inside the average over the random
phase $c$, and we can approximate the kernel using the scalar random
Fourier features
\begin{equation}
z_r(\vx)
:= \sqrt{\tfrac{2}{m}}\cos(\boldsymbol\omega_r^{\!\top}\vx + c_r),
\qquad
k(\vx,\vy)\approx z(\vx)^{\!\top}z(\vy),
\end{equation}
with $\{(\boldsymbol\omega_r,c_r)\}_{r=1}^m$ drawn i.i.d.\ from
$p(\boldsymbol\omega)\times\mathrm{Unif}[0,2\pi]$.
Drawing i.i.d.\ pairs $\{(\boldsymbol\omega_r,c_r)\}_{r=1}^m$ and defining
\begin{equation}
z_r(\vx)
:=
\sqrt{\tfrac{2}{m}}\cos(\boldsymbol\omega_r^{\!\top}\vx + c_r),
\qquad
z(\vx)
:=
\big(z_1(\vx),\ldots,z_m(\vx)\big)^{\!\top},
\end{equation}
we obtain the Monte Carlo approximation
$k(\vx,\vy)\approx \vz(\vx)^{\!\top}\vz(\vy)$.

Applying the augmented Stein operator to each scalar feature yields
\begin{equation}
g_r(\vx;\vtheta)
:=
\big(\mathcal B_{\vtheta,t_0} z_r\big)(\vx),
\qquad
\vg(\vx;\vtheta)
:=
\big(g_1(\vx;\vtheta),\ldots,g_m(\vx;\vtheta)\big)^{\!\top}\in\mathbb R^m.
\label{eq:stein-features-general}
\end{equation}
Substituting $k(\vx,\vy)\approx \vz(\vx)^{\!\top}\vz(\vy)$ into
\eqref{eq:stein-kernel-general} and expanding shows that, up to a
constant factor, the KSD is approximated by the squared norm of the
mean Stein feature,
\begin{equation}
\widehat{\mathrm{KSD}}_{\mathrm{RFF}}^{\,2}(\vtheta)
:=
\bigg\|
\frac{1}{N}\sum_{i=1}^N \vg(\vx_i;\vtheta)
\bigg\|_2^2,
\label{eq:lin-ksd-general}
\end{equation}
which has $O(Nm)$ time complexity. In the
stationary constant–diffusion case, where
$\mathcal B_{\vtheta,t_0}=\mathcal A^{(D)}_{\vtheta}$, this reduces
to the linear–complexity KSD objective used in Eq.~(19) of the main text.

Gradients with respect to $\vtheta$ only require derivatives of the
drift $\vf_{\vtheta}$ (the random features
$\{\boldsymbol\omega_r,b_r\}$ are fixed once sampled), so
\eqref{eq:lin-ksd-general} is well suited to mini–batch stochastic
optimization in the general non–stationary, state–dependent diffusion
setting.

\section{Gene--regulatory network example and KSD-based recovery}
\label{apn:4}

In this section, we detail the gene--regulatory network (GRN) example used in the main text (Sec.~\ref{subsec:gene}) and describe how we recover its parameters from unordered steady-state data using the global KSD route.

\subsection{Seven-dimensional stochastic GRN model}

We construct a $7$-dimensional SDE that couples a three-gene regulatory core to two autonomous ``driver'' oscillators. The state is
\[
  \vx = (p,b,r,x_1,y_1,x_2,y_2)^\top \in \mathbb{R}^7,
\]
where $p,b,r$ are the expression levels of three non-driver genes, and $(x_1,y_1)$ and $(x_2,y_2)$ generate two oscillatory driver signals that modulate the dynamics of $b$ and $r$.

The four driver coordinates follow two noisy limit cycles with slightly different angular frequencies,
\begin{align}
  \mathrm d x_1 &= \Big(\tfrac{\pi}{100} y_1 + x_1(1 - x_1^2 - y_1^2)\Big)\,\mathrm dt
                  + \sqrt{0.0002}\,\mathrm d w_1, \\
  \mathrm d y_1 &= \Big(-\tfrac{\pi}{100} x_1 + y_1(1 - x_1^2 - y_1^2)\Big)\,\mathrm dt
                  + \sqrt{0.0002}\,\mathrm d w_2, \\
  \mathrm d x_2 &= \Big(\tfrac{\sqrt{2}\,\pi}{100} y_2 + x_2(1 - x_2^2 - y_2^2)\Big)\,\mathrm dt
                  + \sqrt{0.0002}\,\mathrm d w_3, \\
  \mathrm d y_2 &= \Big(-\tfrac{\sqrt{2}\,\pi}{100} x_2 + y_2(1 - x_2^2 - y_2^2)\Big)\,\mathrm dt
                  + \sqrt{0.0002}\,\mathrm d w_4.
\end{align}
These terms correspond to two noisy Stuart--Landau oscillators with base frequencies $\pi/100$ and $\sqrt{2}\,\pi/100$ and small isotropic diffusion $D = 10^{-4}$ in each coordinate (the SDE is implemented as
$\mathrm d\vx = f(\vx)\,\mathrm dt + \sqrt{2D}\,\mathrm d\boldsymbol w$,
so $\sqrt{2D} = \sqrt{0.0002}$).

The three non-driver genes use the standard saturating nonlinearity
\begin{align}
  \pi &= \frac{1}{1 + p^2}, &
  \beta &= \frac{1}{1 + b^2}, &
  \rho &= \frac{1}{1 + r^2}.
\end{align}
The two driver inputs BCR and CD40 are smooth functions of the oscillator phases and the current value of $b$, in a way that is qualitatively consistent with the phase-dependent modulation observed in B-cell signaling~\citep{martinez2012quantitative}:
\begin{align}
  \mathrm{BCR}   &= 10\;\big(\sin(\alpha_1)^{30}\big)\,\beta, \\
  \mathrm{CD40}  &= 5\;\big(\sin(\alpha_2)^{30}\big)\,\beta.
\end{align}
Here, $\alpha_1$ and $\alpha_2$ denote the polar angles of $(x_1,y_1)$ and $(x_2,y_2)$, respectively. The exponent $30$ makes the drivers sharply phase-selective while remaining smooth.

We parameterize the regulation of $p,b,r$ by seven nonlinear basis functions of $(p,b,r)$,
\[
  \boldsymbol\phi(p,b,r)
  :=
  \big(\pi,\;\beta,\;\rho,\;\pi\beta,\;\pi\rho,\;\beta\rho,\;\pi\beta\rho\big)^\top \in \mathbb{R}^7,
\]
and a $3\times 7$ interaction matrix $W$ acting on these basis functions. Writing
\(
  \boldsymbol i = (i_1,i_2,i_3)^\top = W\,\boldsymbol\phi
\),
we obtain
\begin{equation}
\begin{bmatrix}
i_1 \\
i_2 \\
i_3
\end{bmatrix}
=
\begin{bmatrix}
0 & 1 & 1 & 0 & 0 & 0 & 0 \\
0 & 0 & 0 & 0 & 0 & 0 & 1 \\
0 & 0 & 1 & 0 & 0 & 0 & 0
\end{bmatrix}
\begin{bmatrix}
\pi \\
\beta \\
\rho \\
\pi\beta \\
\pi\rho \\
\beta\rho \\
\pi\beta\rho
\end{bmatrix},
\end{equation}
which serves as the ground-truth interaction matrix $W^*$.

The SDEs for $(p,b,r)$ read as follows:
\begin{align}
  \mathrm d p &= \Big(10^{-6} - p + 9\,i_1\Big)\,\mathrm dt
                + \sqrt{0.0002}\,\mathrm d w_5, \\
  \mathrm d b &= \Big(2 - (1 + \mathrm{BCR})\,b + 100\,i_2\Big)\,\mathrm dt
                + \sqrt{0.0002}\,\mathrm d w_6, \\
  \mathrm d r &= \Big(0.1 - r + \mathrm{CD40} + 2.6\,i_3\Big)\,\mathrm dt
                + \sqrt{0.0002}\,\mathrm d w_7.
\end{align}
Collecting all terms, the full $7$-dimensional drift can be written as
\[
  f_{\vtheta}(\vx)
  = U(\vx)\,\vtheta + v(\vx),
\]
where $\vtheta\in\mathbb{R}^{21}$ stacks the entries of $W$ row-wise, $U(\vx)$ encodes the basis functions for each of $p,b,r$, and $v(\vx)$ contains the parameter-free parts (the driver oscillators and constant terms).

Our goal in this example is to infer the $3\times 7$ matrix $W$ (equivalently, the $21$-dimensional vector $\vtheta$) from unordered $7$-dimensional snapshots drawn from the stationary distribution of the SDE.

\subsection{Simulation of the data}

We simulate the SDE with the true parameter vector
\[
  \vtheta^* = (0,1,1,0,0,0,0,\;0,0,0,0,0,0,1,\;0,0,1,0,0,0,0)^\top,
\]
which implements the interaction matrix $W^*$ above. The diffusion matrix is constant and diagonal,
\[
  D = 10^{-4} I_7,
\]
so the noise amplitude is $\sqrt{2D} = \sqrt{0.0002}$ in each coordinate.

We integrate the SDE using a deterministic fourth-order Runge-Kutta (RK4) step for the drift plus an Euler-Maruyama step for the noise:
\[
  \vx_{t+\Delta t}
  =
  \vx_t + \frac{\Delta t}{6}(s_1 + 2s_2 + 2s_3 + s_4)
          + \sqrt{\Delta t}\,\boldsymbol\eta_t\sqrt{2D},
\]
where $s_k$ are the usual RK4 stages evaluated with the drift $f_{\vtheta^*}$, and $\boldsymbol\eta_t\sim\mathcal N(0,I_7)$.

To obtain approximately i.i.d.\ cross-sectional samples from the stationary distribution, we proceed as follows:
\begin{itemize}
  \item Initialize a mini-batch of states $x\in\mathbb{R}^{B\times 7}$ with $B=2048$, sampling all coordinates from a standard normal, and enforcing $p,b,r>0$ by taking their absolute values.
  \item Run a burn-in phase of $30{,}000$ time steps with $\Delta t = 10^{-2}$ using the RK4+noise step above.
  \item After burn-in, continue simulating and record every $10$-th step (thinning factor $10$) until we have collected $N=50{,}000$ samples of $\vx=(p,b,r,x_1,y_1,x_2,y_2)$.
\end{itemize}
This procedure yields a dataset $X\in\mathbb{R}^{N\times 7}$ of unordered steady-state samples, which we then use as input to the global KSD route.

\subsection{Global KSD estimator and linear system in \texorpdfstring{$\vtheta$}{theta}}

Because the drift is affine in $\vtheta$ and the diffusion matrix is constant, the Stein features arising from our KSD construction are linear in $\vtheta$. We approximate the Gaussian RBF kernel using $m=2048$ random Fourier features with bandwidth $\ell$ chosen by the median heuristic (\cite{garreau2017large})  on the simulated data. For each random frequency $\boldsymbol\omega_r\sim\mathcal N(0,\ell^{-2}I_7)$ and phase $c_r\sim\mathrm{Unif}[0,2\pi]$ we define
\[
  z_r(\vx) = \sqrt{\tfrac{2}{m}}\cos(\boldsymbol\omega_r^{\!\top}\vx + c_r),
\]
and apply the diffusion--Stein operator to each scalar feature, obtaining
\[
  g_r(\vx;\vtheta)
  = \big(\mathcal A^{(D)}_{\vtheta} z_r\big)(\vx),
  \qquad r=1,\dots,m.
\]
For our affine-in-parameter GRN drift, each $g_r(\vx;\vtheta)$ can be written as
\[
  g_r(\vx;\vtheta)
  = a_r(\vx)^\top \vtheta + \nu_r(\vx),
\]
where $a_r(\vx)\in\mathbb{R}^{21}$ depends only on the basis functions
$\boldsymbol\phi(p,b,r)$ and the projection of $\boldsymbol\omega_r$ onto $(p,b,r)$, and $\nu_r(\vx)$ collects all parameter-free terms (including the driver oscillators and the diffusion contribution).

Averaging over the data gives the empirical mean Stein feature
\[
  \widehat{\boldsymbol g}(\vtheta)
  :=
  \frac{1}{N}\sum_{i=1}^N \vg(\vx_i;\vtheta)
  =
  A_{\mathrm{global}}\,\vtheta + \vb_{\mathrm{global}},
\]
where $\vg(\vx;\vtheta)=(g_1(\vx;\vtheta),\dots,g_m(\vx;\vtheta))^\top\in\mathbb{R}^m$ and
\[
  A_{\mathrm{global}} \in \mathbb{R}^{m\times 21},
  \qquad
  \vb_{\mathrm{global}} \in \mathbb{R}^m
\]
are computed in closed form by batched accumulation over the dataset (batch is used only when sample size is too large). The linear-time RFF approximation to the KSD is then
\[
  \widehat{\mathrm{KSD}}^2_{\mathrm{RFF}}(\vtheta)
  = \big\|\widehat{\boldsymbol g}(\vtheta)\big\|_2^2
  = \|A_{\mathrm{global}}\vtheta + \vb_{\mathrm{global}}\|_2^2.
\]
Minimizing this quadratic objective is equivalent to solving a regularized linear system. We use Tikhonov regularization with $\lambda=10^{-6}$ and calculate the $\widehat{\vtheta}\in\mathbb{R}^{21}$.

Across $50$ independent simulation and estimation runs (different random seeds and data draws), the recovered interaction matrices $\widehat{W}$ closely match $W^*$ for most entries, with one coefficient systematically deviating. As discussed in the main text, this misestimate entry lies in a weakly constrained direction of parameter space predicted by our Gram/Hessian analysis, and the corresponding SDE trajectories in $(p,b,r)$ remain practically indistinguishable from those generated by the true parameters.

\section{Density estimator for ``dynamics-to-density''}

\begin{figure*}[ht]
\vskip 0.2in
\begin{center}
\centerline{\includegraphics[width=0.99\columnwidth]{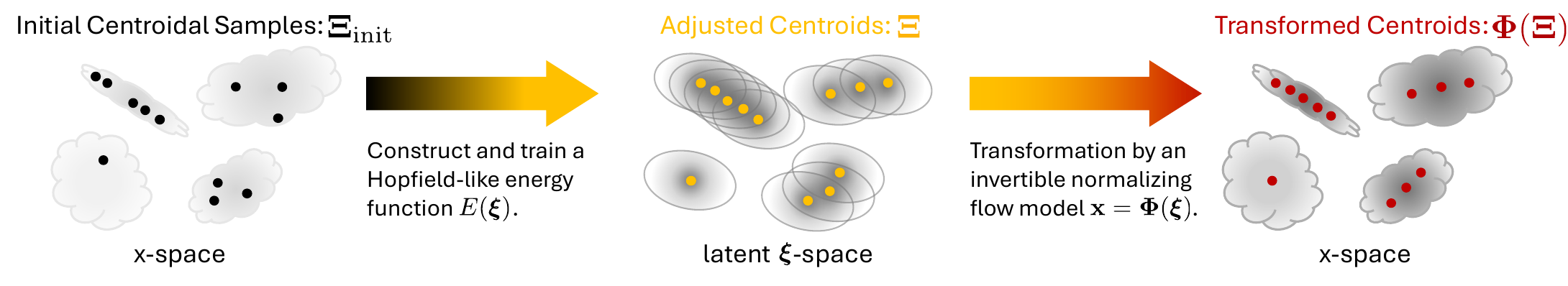}}
\caption{Schematic of our density estimator, comprising a GMM in a latent space $\boldsymbol{\xi}$ (also functioning as a Hopfield-like energy) with trainable centroids $\boldsymbol{\Xi}$, and a RealNVP normalizing flow $\boldsymbol{\Phi}$. The latent GMM provides a flexible approximation to the target density while reducing artifacts in sparse regions; the flow, initialized near the identity, refines this approximation to match the desired density more accurately.}
\label{fig:fig4}
\end{center}
\vskip -0.2in
\end{figure*}

We propose a versatile density estimator with \emph{direct} access to both the normalized density \(p(\boldsymbol{x})\) and the score function \(\mathbf{s}(\boldsymbol{x})=\nabla_{\boldsymbol{x}}\log p(\boldsymbol{x})\). The score is required to train the estimator from dynamics via the score-based Fokker–Planck residual.

The main idea is to use a latent GMM as the backbone, augmented with a normalizing flow layer (a near-identity RealNVP model~\cite{dinh2016density}) for additional flexibility, as illustrated in Fig.~\ref{fig:fig4}. The latent distribution of $\boldsymbol{\xi}$ is a trainable GMM with $n$ centroids arranged as a $d\times n$ matrix $\boldsymbol{\Xi}$ and a shared covariance matrix $\boldsymbol{\Sigma}$. The log-weights \(\boldsymbol{\lambda}\in\mathbb{R}^n\) of the centroids are also trainable. Unlike a VAE, the latent space has the same dimension as the data space. A near-identity RealNVP flow $\boldsymbol{\Phi}$ then warps the latent GMM to better match the desired data density.

Up to an additive constant, the latent energy is
\begin{align}
E_{\xi}(\boldsymbol{\xi})
&=
\frac{1}{2}\,\boldsymbol{\xi}^{\!\top}\boldsymbol{\Sigma}^{-1}\boldsymbol{\xi}
-
\operatorname{lse}\!\Big(\boldsymbol{\Xi}^{\!\top}\boldsymbol{\Sigma}^{-1}\boldsymbol{\xi}
+ \boldsymbol{\lambda} + \boldsymbol{l}\Big), \label{eq:latent-energy}\\[-2pt]
[\boldsymbol{l}]_i &:= -\tfrac{1}{2}\,[\boldsymbol{\Xi}]_{:,i}^{\!\top}\boldsymbol{\Sigma}^{-1}[\boldsymbol{\Xi}]_{:,i}.
\end{align}
Thus \(p_{\xi}(\boldsymbol{\xi}) \propto \exp\{-E_{\xi}(\boldsymbol{\xi})\}\).

\paragraph{Density, energy, and score in data space.}
With \(\boldsymbol{\xi}=\boldsymbol{\Phi}^{-1}(\boldsymbol{x})\) and Jacobian
\(\mathbf{J}_{\xi|x}(\boldsymbol{x}):=\partial \boldsymbol{\Phi}^{-1}(\boldsymbol{x})/\partial \boldsymbol{x}\),
which is easily computed for RealNVP, we have
\begin{align}
p(\boldsymbol{x})
&= p_{\xi}\!\big(\boldsymbol{\Phi}^{-1}(\boldsymbol{x})\big)\,
\Bigl|\det \mathbf{J}_{\xi|x}(\boldsymbol{x})\Bigr|, \label{eq:px}\\
E(\boldsymbol{x})
&= -\log p(\boldsymbol{x})
= E_{\xi}\!\big(\boldsymbol{\Phi}^{-1}(\boldsymbol{x})\big)
- \log\Bigl|\det \mathbf{J}_{\xi|x}(\boldsymbol{x})\Bigr|
+\text{const.} \label{eq:Ex}
\end{align}
Beyond the density and energy, which are relevant for likelihood-based training, we also require the score to train this model \emph{from dynamics alone} using the Fokker–Planck residual.

Define
\[
\boldsymbol{\alpha}(\boldsymbol{\xi})
:= \operatorname{softmax}\!\Big(\boldsymbol{\Xi}^{\!\top}\boldsymbol{\Sigma}^{-1}\boldsymbol{\xi}
+ \boldsymbol{\lambda} + \boldsymbol{l}\Big),
\]
then the latent score is
\begin{equation}
\mathbf{s}_{\xi}(\boldsymbol{\xi})
:= \nabla_{\boldsymbol{\xi}}\log p_{\xi}(\boldsymbol{\xi})
= \boldsymbol{\Sigma}^{-1}\!\big(\boldsymbol{\Xi}\,\boldsymbol{\alpha}(\boldsymbol{\xi}) - \boldsymbol{\xi}\big).
\label{eq:sxi}
\end{equation}
By the chain rule, the data-space score is
\begin{equation}
\mathbf{s}(\boldsymbol{x})
:= \nabla_{\boldsymbol{x}}\log p(\boldsymbol{x})
= \mathbf{J}_{\xi|x}(\boldsymbol{x})^{\!\top}\,\mathbf{s}_{\xi}\!\big(\boldsymbol{\Phi}^{-1}(\boldsymbol{x})\big)
+ \nabla_{\boldsymbol{x}}\log\Bigl|\det \mathbf{J}_{\xi|x}(\boldsymbol{x})\Bigr|.
\label{eq:sx}
\end{equation}

\paragraph{Training from dynamics (no data).}
Given a known SDE drift \(\mathbf{f}(\boldsymbol{x})\) and diffusion \(\boldsymbol{G}\), we learn the steady-state density by minimizing a sum of squared steady Fokker–Planck residuals, using \(\mathbf{s}(\boldsymbol{x})\) from \eqref{eq:sx}. Only the centroids \(\boldsymbol{\Xi}\), mixture weights \(\boldsymbol{\lambda}\), shared covariance \(\boldsymbol{\Sigma}\), and the parameters of the near-identity flow \(\boldsymbol{\Phi}\) are optimized.

\paragraph{2D SDE demonstration.}
In our 2D example, we start from \(n=10\) initial centroids and train the estimator to match the true steady-state density of a planar SDE. During training, the centroids move to cover the high-density regions while the (small) nonlinear flow refines the shape of the level sets, resulting in close agreement between the learned and true densities, as shown in the main text (Fig.~\ref{fig:dyn2dens}).

\end{document}